\documentstyle[prl,aps,floats,epsf]{revtex}

\title{The exact solutions of differential equation with delay} 
\author{K. Hasebe}
\address{Faculty of Business Administration,
Aichi University, Miyoshi, Aichi 470-0296, Japan \\
e-mail address: hasebe@aichi-u.ac.jp}
\author{A. Nakayama}
\address{Gifu Keizai University, Ohgaki, Gifu 503-8550, Japan \\
e-mail address: g44153g@nucc.cc.nagoya-u.ac.jp}
\author{Y. Sugiyama}
\address{Division of Mathematical Science, 
City College of Mie, Tsu, Mie 514-0112, Japan \\
e-mail address: genbey@eken.phys.nagoya-u.ac.jp} 
\date{December 1, 1998}

\begin{document}
\twocolumn[\hsize\textwidth\columnwidth\hsize\csname
@twocolumnfalse\endcsname
\maketitle
\begin{abstract}
The exact solutions of the first order differential equation
with delay are derived. The equation has been introduced as
a model of traffic flow. The solution describes the traveling cluster 
of jam, which is
characterized by Jacobi's elliptic function. We also obtain
the family of solutions of such type of equations.
\end{abstract}

\hspace*{1.8cm}

\vspace{1pt}
\hspace*{1.9cm}{patt-sol/9812003}

\vskip1.9pc]
\narrowtext  


In this paper, we investigate  following differential-difference
equation,
\begin{equation}
\frac{d}{dt}{x}_n(t + \tau)=f( x_{n+1}(t)-x_n(t) ),
\label{eqx}
\end{equation}
where $\tau$ is a real positive constant
called ``delay''.
The index $n$ takes integer.
The set of equations of this type has been introduced 
for a car-following model
of traffic flow \cite{Newell}\cite{Whitham}.
In that case $x_n$ is the position of the $n$th car.
Equation (\ref{eqx}) has been popular in many physical phenomena 
of relaxation towards an optimal equilibrium state,
such as relaxation effect in gases, chemical reactions and
synchronization problem. 

As a model for traffic flow, 
$f(x)=\tanh (x)$ is the reasonable choice \cite{ov}\cite{hbnns}, 
which was first introduced in the
model \cite{ov},
\begin{equation}
\tau\frac{d^2}{dt^2}{x}_n(t)= f(x_{n+1}(t)-x_n(t))-\frac{d}{dt}{x}_n(t).
\label{ovm}
\end{equation}
This model may have some relation to Eq.(\ref{eqx}) for relatively small $\tau$
\cite{Whitham}.
Many studies have been made to the model of Eq.(\ref{ovm}) with 
$f(x)=\tanh(x)$, which has the stable traveling
cluster solution as traffic jam \cite{ov}.
The traveling cluster solution is characterized as a soliton of 
modified Korteweg-de Vries (MKdV)
equation in the vicinity of the critical point \cite{KS}.
One interesting question is whether Eq.(\ref{eqx}) has such
traveling cluster solution or not.
Our simulation result suggest the existence of such solution,
which describes a traveling cluster moving backward
with the velocity $v=1/(2\tau)$ \cite{hbnns}.
Recently, we have an information that 
Igarashi, Itoh and Nakanishi have found the exact solution
of Eq.(\ref{eqx}), which characterized by the theta
function \cite{private}.
In this paper we also derive a series of analytic solutions
of traveling cluster in the context of our work, 
and confirm their stability by the numerical simulations.

It is convenient to introduce new variable $h_n(t) = x_{n+1}(t)-x_n(t)$
and rewrite Eq.(\ref{eqx}),
\begin{equation}
\frac{d}{dt}{h}_n(t + \tau)= f( h_{n+1}(t) ) - f( h_n(t) ) \ .
\label{eqh}
\end{equation}
We start at the linear theory  
assuming that the amplitude $h_n(t)$ is infinitesimal and
$f(h_n)=f(0)+f'(0)h_n$. Equation (\ref{eqh}) becomes 
\begin{equation}
\frac{d}{dt} h_n(t + \tau)=  h_{n+1}(t) -  h_n(t)\ .
\label{linear_eq}
\end{equation}
Here we set $f'(0)=1$ without loss of generality.
We investigate the solution of the form
\begin{equation}
h_n(t) = \exp (i\alpha n + i\omega t) \ 
\label{linear_sol}
\end{equation}
for real $\alpha$ and $\omega$.
Inserting Eq. (\ref{linear_sol}) to Eq. (\ref{linear_eq}) we obtain
\begin{equation}
\frac{\sin \alpha/2}{\alpha/2} = \frac{1}{2\tau},
\label{omega_eq}
\end{equation}
and
\begin{equation}
\alpha=2\omega \tau \ .
\label{alpha_eq}
\end{equation}
Equation (\ref{omega_eq}) has a solution if $2\tau$ is larger than $1$,
which means
$\tau = 1/2$ is critical.
If $\tau$ is a little bit larger than $1/2$,
Eq.(\ref{omega_eq}) has only two solutions $\alpha =\pm \lambda$.
Then, we obtain the solution of Eq.(\ref{linear_eq}) as
\begin{equation}
h_n(t) = \exp \pm i\lambda (n+\frac{t}{2\tau}) \ .
\end{equation}
This represents a traveling wave solution with the velocity
$1/(2\tau)$
in the space of index $n$, which is treated as  
a continuous variable. The wave moves backward against the numbering direction,
which appears as the traveling wave in the real space moving backward
in the flow of $x_n$.
The above analysis is first given by Whitham \cite{Whitham}.

Now let us investigate the exact traveling wave solution of Eq.(\ref{eqh}). 
We treat the index $n$ as a continuous variable, and change the notation
$h_n(t)$ to $h(n,t)$.
We introduce 
 new variables on the moving flame of traveling wave as
$u=n+vt$, where
$v$ is the velocity of the traveling wave.
We search the solution which does not change its form on this flame. 
We define the amplitude of traveling wave
\begin{equation}
H(u) =  H(n+vt) \equiv h(n,t) .
\end{equation}
Eq.(\ref{eqh}) for the amplitude $H(u)$ is expressed as
\begin{equation}
v\frac{d}{du}H(u+v\tau)
= f(H(u+1))-f(H(u)) .
\end{equation}
Replacing $u$ by $u-1/2$,
we get more symmetric form 
\begin{equation}
v\frac{d}{du}
H(u+\sigma)
= f(H(u+\frac{1}{2}))-f(H(u-\frac{1}{2})) ,
\label{static_eq}
\end{equation}
where
\begin{equation}
\sigma = v\tau-\frac{1}{2}.
\end{equation}
We investigate the solution under the condition $\sigma = 0$.
This means that the traveling wave of such solutions
propagates backward with just the same velocity as the linear theory,
\begin{equation}
v=\frac{1}{2\tau}.
\end{equation}

Now we present the definite form of exact solutions giving concrete
 examples of $f(x)$.
First we take  $f(x) = \tanh(x)$, which is a suitable choice for a model 
 of traffic flow.
Introducing a new amplitude
\begin{equation}
G=f(H),
\label{new_var}
\end{equation}
Eq.(\ref{static_eq}) with $\sigma = 0$ is rewritten as
\begin{equation}
v\frac{dG(u)/du}{1-G(u)^2}
=G(u+\frac{1}{2})-G(u-\frac{1}{2}).
\label{G-_eq}
\end{equation}
We can easily find a solution of Eq.(\ref{G-_eq}) in the form
\begin{equation}
G(u) = \beta \;\begingroup\rm{sn}\endgroup (\alpha u,k),
\label{sn_solution}
\end{equation}
where sn is Jacobi's elliptic function with modulus $k$.
The parameter $\alpha$ is determined by
\begin{equation}
\frac{ \begingroup\rm{sn}\endgroup (\alpha/2,k)}{\alpha/2}=\frac{1}{2\tau},
\label{alpha_eq1}
\end{equation} 
and $\beta$ is given by
\begin{equation}
\beta= \pm  k \frac{\alpha}{4\tau} .
\label{beta_eq1}
\end{equation}
Eq.(\ref{alpha_eq1}) has a real solution only if
$1/(2\tau)< 1$.
In the case of $k=0$, Eq.(\ref{alpha_eq1}) reduces to the result of
linear theory, Eq.(\ref{omega_eq}).
The modulus $k$ is a free parameter of the solution,
which indicates the existence of many
solutions for the same traveling velocity $v = 1/(2\tau)$.
The relation between the modulus and solutions is discussed later.
Next we take $f(x) = \tan(x)$. In this case Eq.(\ref{G-_eq})
is replaced by
\begin{equation}
v\frac{dG(u)/du}{1+G(u)^2}
=G(u+\frac{1}{2})-G(u-\frac{1}{2}).
\label{G+_eq} 
\end{equation}
It is also  easy to see that 
Eq.(\ref{G+_eq})
has a solution
\begin{equation}{\displaystyle
G(u) = \pm k \frac{\alpha}{4\tau}\;\begingroup\rm{cn}\endgroup (\alpha u,k),}
\label{cn_solution}
\end{equation}
where cn is another Jacobi's elliptic function. The parameter $\alpha$ is
determined by
\begin{equation}
\frac{ \begingroup\rm{sd}\endgroup (\alpha/2,k)}{\alpha/2}=\frac{1}{2\tau},
\label{alpha_eq2}
\end{equation} 
which also reduces to the result of linear theory, if we take
 $k=0$.

The above two solutions suggest us to expect the existence of 
another solutions using elliptic functions. Actually, we have found the
family of solutions of Eqs. (\ref{G-_eq}) or (\ref{G+_eq}).
The solutions of this family are denoted in the form
\begin{equation}
G(u) = \pm C_k \frac{\alpha}{4\tau}\;\begingroup\rm{g}\endgroup (\alpha u,k),
\label{g_solution}
\end{equation}
where $\alpha$ is
determined by
\begin{equation}
\frac{ \begingroup A\endgroup (\alpha/2,k)}{\alpha/2}=\frac{1}{2\tau}.
\label{alpha_eq3}
\end{equation} 
In the above, g and $A$ are appropriate elliptic functions.
The solution is given by each set of ($\rm{g}$,$A$,$C_k$) 
in Tables \ref{tbl:tanh} and \ref{tbl:tan}. 

\begin{table}[htb]
\caption{Solutions for $f(x) = \tanh(x)$
\label{tbl:tanh}}
\vspace{4pt}
\hspace{2cm}
\begin{tabular}{|c||c|c|c|c|c|c|c|c|} \hline
$\rm{g}$ & sn  & ns  & sc             & cs  & ds  & cd  & dc  & nc \\ 
\hline
$A$ & sn  & sn  & sc             & sc  & sd  & sn  & sn  & sd \\ 
$C_k$    & $k$ & $1$ & $\sqrt{1-k^2}$ & $1$ & $1$ & $k$ & $1$ & $\sqrt{1-k^2}$
 \\ \hline
\end{tabular}
\end{table}
\begin{table}[htb]
\caption{Solutions for $f(x) = \tan(x)$
\label{tbl:tan}}
\vspace{4pt}
\hspace{3.5cm}
\begin{tabular}{|c||c|c|c|c|} \hline
$\rm{g}$ & cn & nd & dn & sd \\ 
\hline
$A$ & sd & sc & sc  & sd \\ 
$C_k$    & $k$ & $\sqrt{1-k^2}$ & $1$ & $k\sqrt{1-k^2}$
 \\ \hline
\end{tabular}
\end{table}

\begin{figure}[htb]
\epsfxsize = 8cm
\hfil\epsfbox{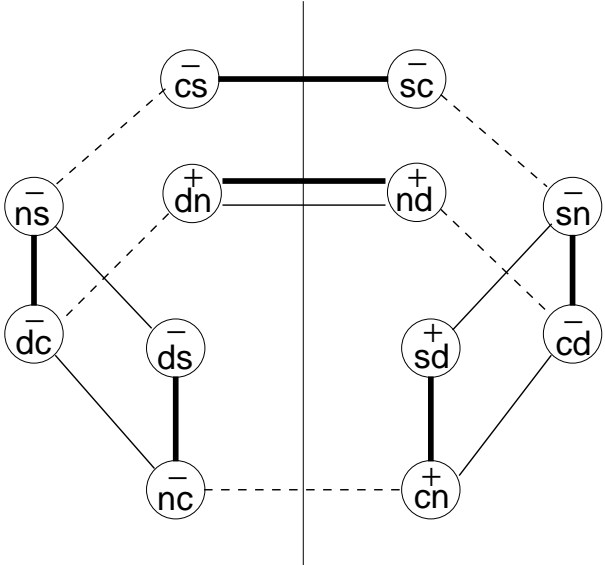}\hfill
\caption{The relation of elliptic solutions for 
Eq. (\ref{G-_eq}) denoted by $-$ sign and Eq. (\ref{G+_eq}) 
denoted by $+$ sign.
They are connected by transformations $k \rightarrow ik$, 
$\alpha \rightarrow i\alpha$ and $u \rightarrow u+K(k)/\alpha$, drawn
by thin lines, dashed lines and thick lines, respectively.
They form a double hexagonal structure.
Each pair of two elliptic functions connected with a thick line gives
an equivalent solution.
The functions in the left and right sides 
divided by the vertical center line are inverse with each other.
\label{fig:hexisa}}
\end{figure}

Thus, all Jacobi's elliptic functions are the solutions for either Eq. 
(\ref{G-_eq}) or (\ref{G+_eq}).
This result is well understood from the following fact.
Jacobi's elliptic functions $\begingroup\rm{g}\endgroup (\alpha u,k)$ are
connected each other by Jacobi's imaginary transformation:
$\alpha \rightarrow i\alpha$, the imaginary
transformation of modulus: $k \rightarrow ik$ and the translation of
$u$: $u \rightarrow u+K(k)/\alpha$, where $K(k)$ is a quarter of the
period of elliptic functions.
These transformations preserve the form of Eq. 
(\ref{G-_eq}) or (\ref{G+_eq}), or exchange each other.
Each pair of two elliptic functions connected by the translation gives
an equivalent solution.
The relation of solutions is shown in Fig.\ref{fig:hexisa}.
%


The explicit form for the solution of Eq.(\ref{eqh}) is given by $H=f^{-1}(G)$.
In this step some solutions for $G$ needs appropriate interpretation
for the realistic meaning. 
The solution $G\sim$ cs for $G=\tanh(H)$ is the case, which
has the region of $u$ where $|G|>1$. On the other hand,
the solution $G\sim$ sn has a realistic meaning in the whole range
of $u$. 
 
Equation (\ref{G-_eq}) or (\ref{G+_eq}) includes the solutions
for another choice of $f(x)$ besides $\tanh(x)$ and $\tan(x)$.
Actually, $f(x)=\coth(x)$ leads to Eq.(\ref{G-_eq}),
and $f(x)=\cot(x)$ leads to Eq.(\ref{G+_eq}) with 
$v\rightarrow -v$. Jacobi's elliptic functions also provide the solutions
for these systems. 
This result presents some interesting picture. Let us compare the 
systems controlled by $f(x)=\tanh(x)$ and $\coth(x)$.
At first sight, the behavior of these two systems seems different from
each other. We note  sn and ns are the solutions for both systems,
which obey Eq.(\ref{G-_eq}).
The solution $H \sim \rm{arccoth(\rm{ns})}$ 
in the system of $f(x)=\coth(x)$ has essentially the same form as the solution
$H \sim \rm{arctanh(\rm{sn})}$ in the system of $f(x)=\tanh(x)$. 
As the result, these two systems may
have the same global phenomenon in spite of the different type 
of local interaction. 
This analysis is not simply applicable to the solution nc in the 
system of $f(x)=\coth(x)$, because cn is the solution of the system
of $f(x)=\tan(x)$, which obeys the other Eq.(\ref{G+_eq}). 
 
We discuss the meaning of the modulus in our solution.
The modulus $k$ determines the period of elliptic functions, which
 is related to the number of traveling cluster and the
 boundary condition.
For example, $k = 1$ gives kink like solution corresponding to the
boundary condition $G(-\infty) = -G(\infty)$ ,
\begin{equation}
G(z)=\tanh(\alpha/2)\tanh(\alpha u)
\end{equation} 
with $\alpha$ satisfying
\begin{equation}
\frac{\tanh(\alpha / 2)}{\alpha /2}=\frac{1}{2\tau}.
\end{equation}

\begin{figure}[htb]
\epsfxsize = 10cm
\hfil\epsfbox{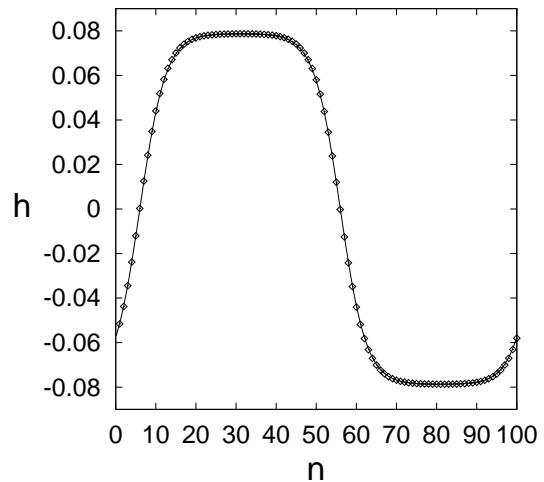}\hfill
\caption{Simulation result of $h_n$ in the system of $f(x)=\tanh(x)$
is shown by diamonds together with the  analytic solution
given by sn. The curve of analytic solution is shifted 
by $u \rightarrow u+$constant.
\label{fig:sim_sn}}
\end{figure}

We perform the simulation to check the stability of our analytic
solution. 
Fig.\ref{fig:sim_sn} shows an
elliptic solution for $f(x)=\tanh(x)$ given by sn 
\begin{equation}
H(u,k) = \begingroup\rm{arc}\endgroup \tanh(k \frac{\alpha}{4\tau}\;
\begingroup\rm{sn}\endgroup (\alpha u,k)),
\label{solution_sim}
\end{equation}
with $\tau=0.501$, $k=0.9965$ and $\alpha=0.15522$ 
together with the result of simulation for Eq.(\ref{eqx}) performed 
in the periodic boundary condition with a suitable initial condition.

%

We are convinced the set of Eqs. (\ref{G-_eq}) and (\ref{G+_eq}) has much
more solutions constructed with elliptic functions.
As a fact, 
the solution found by Igarashi, Itoh and Nakanishi \cite{IIN} is rewritten
in our formulation as
\begin{equation}
G(u)=\beta \frac{ \begingroup\rm{sn}\endgroup (\alpha(u+a))
            +\begingroup\rm{sn}\endgroup (\alpha(u-a))}
           {\begingroup\rm{sn}\endgroup (\alpha u)} + \gamma ,
\label{asymmetric_sol}
\end{equation}
where $a$ is a free parameter. Actually this satisfies Eq.(\ref{G-_eq}).
Their solution includes
Eq.(\ref{sn_solution}) as a special case
if we take $\alpha a = K(k)/2$ 
\cite{Nakanishi}. 
The parameter $\alpha$ is determined by
\begin{equation}
\frac{\alpha}{4\tau}\frac{{\rm cn}(\alpha a){\rm dn}(\alpha a)} {{\rm sn}(\alpha a)}
+
\frac{x_+x_-}{x_+-x_-}=0,
\end{equation}
where
\begin{equation}
x_\pm = 1-\frac{{\rm sn}^2(\alpha a)}{{\rm sn}^2(\alpha(a\pm \frac{1}{2}))}.
\end{equation}
Then $\beta$ and $\gamma$ are given by
\begin{equation}
\beta=\pm\frac{\alpha}{4\tau {\rm sn}(\alpha a)},
\end{equation}
and
\begin{equation}
\gamma =\pm\frac{x_++x_-}{x_+-x_-}.
\end{equation}
This solution represents the
asymmetry of the widths of upper and lower plateaus for traveling cluster 
in contrast of the solution given by a single sn shown in 
 Fig.\ref{fig:sim_sn}.
Some combinations of elliptic functions in the numerator and denominator 
of Eq.(\ref{asymmetric_sol}) are solutions, 
those are constructed by the three transformations represented in 
Fig. \ref{fig:hexisa}.
This type of solutions has asymmetry in contrast of the solutions given by
a single elliptic function in Tables \ref{tbl:tanh} and \ref{tbl:tan}.
Probably, the set of Eqs. (\ref{G-_eq}) and (\ref{G+_eq}) has some
algebraic structure in the space of  solutions constructed
with elliptic functions.

We remark all these solutions have the common velocity $v = 1/(2\tau)$.
Numerical simulation shows that 
the traveling cluster solutions of Eq.(\ref{eqx}) 
preserve their velocity as $v = 1/(2\tau)$ in the deformation of $f(x)$
beyond $\tanh(x)$.
This fact suggests $v$ is some invariant quantity
of the structure in the set of the solutions.  

The set of differential-difference equations for $G$: 
Eqs. (\ref{G-_eq}) and (\ref{G+_eq})
offers rich contents of the system with traveling cluster solutions
which characterized by elliptic functions.
Eqs. (\ref{G-_eq}) and (\ref{G+_eq}) are related to
some soliton systems.
Our equations can be derived as the traveling wave equations 
for such soliton systems.
Eqs. (\ref{G-_eq}) and (\ref{G+_eq}) correspond to
one of the evolution equations 
discussed by Ablowitz and Ladik \cite{AL}. 
The soliton systems related to Eq.(\ref{G+_eq}) were widely
discussed in the self dual network equations of nonlinear inductors
and capacitors by Wadati \cite{W}, Hirota and Satsuma \cite{HS}.
Wadati showed the corresponding soliton system to Eq.(\ref{G+_eq}) 
was derived from Lotka-Volterra system 
by B\"acklund transformation \cite{W}. 
These soliton systems reduce to MKdV equation.
Our system of difference equation with delay may be related
to soliton systems.
 

This work was partly supported 
by a Grant-in-Aid (No. 10650066) of the Japanese Ministry of Education,
Science, Sports and Culture.



\end{document}